# David Gill - Magnificent and Desirable Astronomer

*John Reid, Department of Physics, Meston Building, The University, Aberdeen AB9 2UE, UK*

*Abstract*

This paper was presented at the International Conference on the History of Physics held at Trinity College, Cambridge, in September 2014. It was given to mark the centenary of the death of David Gill, the foremost British astronomer in the last quarter of the 19th century and into the 20th century. The story of Gill's conversion from a successful career as a clock and watchmaker to one of the most respected astronomers in the world is fascinating in itself. Gill's speciality was in astrometry, an area of astronomy of both practical and scientific importance that tended to be eclipsed in the 20th century by the rise of astrophysics. As Her Majesty's Astronomer at the Cape of Good Hope for 27 years, David Gill was admired for his prolific contribution of highly accurate and trustworthy results. David Gill's collaboration was desired by leading astronomers of the day and he was the only southern hemisphere representative on the hugely important Conférence Internationale des Étoiles Fondamentales of 1896. He created with Jacobus Kapteyn the first extensive star catalogue derived from photographic plates (the CPD), including over 450,000 stars. He was an initiator of the biggest multi-national and multi-observatory project of the century, taking on the rôle of President of the International Commission overseeing the Carte du Ciel project. David Gill raised the Cape of Good Hope observatory from a rather run-down jobbing observatory of the Admiralty to the best observatory in the Southern hemisphere. He acquired in the process the reputation of an expert in precision instrument design. He also made a substantial name for himself as semi-official director of the geodetic survey of South Africa. Overall David Gill was highly influential, respected and well liked. With the resurgence of interest in astrometry, the contribution of David Gill to astronomy is due for a re-assessment a century after he died.

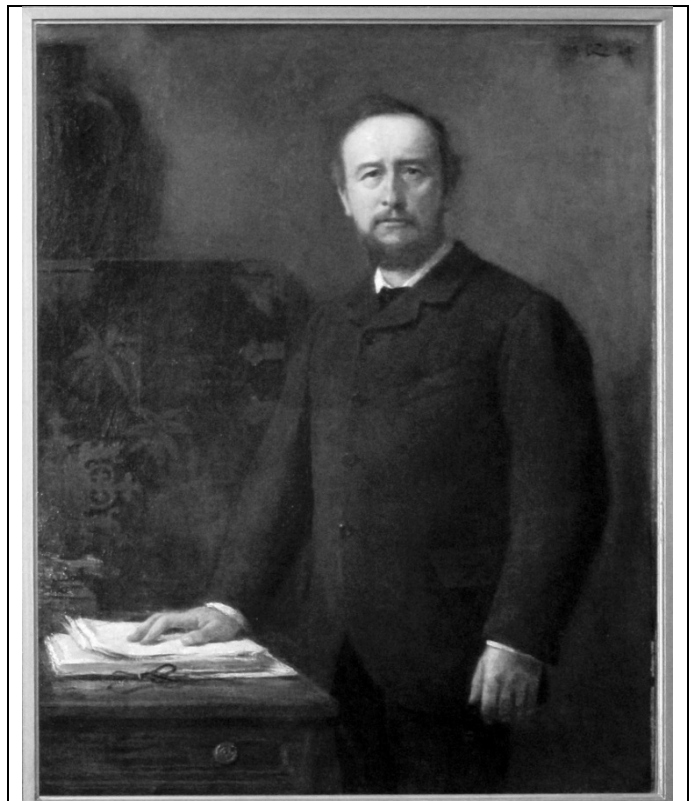

*Fig. 1  David Gill aged about 40 from a portrait by George Reid, PRSA, courtesy Royal Astronomical Society.*

*Introduction*

My title is a nod to James Lequeux's biography *Le Verrier: Magnificent and Detestable Astronomer[1]*, which appeared in translation in 2013. Gill was an astronomer of the generation after Le Verrier. He was magnificent in his achievements and hugely well liked both personally and as a professional astronomer. He is still the only Scot to have received the Watson medal for astronomy of the US National Academy of Sciences and the coveted Bruce medal of the Astronomical Society of the Pacific for life-time achievement in astronomy, and the Gold medal of the Royal Astronomical Society twice. He received the Royal Medal of the Royal Society of London and his portraits in oils hang in both the Royal Society of London and the Royal Astronomical Society. He was their President for two years. He had built up the best observatory in the Southern hemisphere, made notable contributions to astronomy, been knighted by Queen Victoria in 1900 and received honours in at least three continents. Arguably he has been Scotland's most successful astronomer. For many in the present day, though, David Gill is perhaps the most famous astronomer they've never heard of.

*Watchmaker to the Queen*

The story of how Gill got into professional astronomy is fascinating, for in his mid 20s he was a clockmaker, proprietor of a successful watch and clock business, "Watchmakers to the Queen" no less, with a shop in Aberdeen's main thoroughfare. He sold the business when he was 28 in 1872[2].

My story today begins with Gill in 1878, by then an unemployed watchmaker from Aberdeen with no university degree (though I have to say that he had been to the undergraduate classes of James Clerk Maxwell almost 20 years earlier), competing for the post of Her Majesty's Astronomer at the Cape of Good Hope with one other candidate: William Christie, Cambridge wrangler and for 8 years assistant to Airy at the Greenwich Observatory. The post was in the gift of the Lords Commissioners of the Admiralty, presided over by the First Lord of the Admiralty, one W H Smith – the same whose family firm ran the eponymous railway news and book stands that had sprung up everywhere. Smith chose Gill. As it would turn out, the appointment was inspired, for the Admiralty, for astronomy and for South Africa. It may have been some consolation for Christie that he was appointed Astronomer Royal to succeed Airy in 1881 but in spite of his 28 years in this office he wasn't half the astronomer that Gill was.

As an aside I'll mention that 13 years on from his appointment, Gill's reputation was such that he was effectively offered the Lowndean Professorship of Astronomy & Geometry at the University of Cambridge upon the death of John Couch Adams. He turned it down, to his wife's regret. He did, though, accept later on an Honorary Doctorate (ScD) from Cambridge.

*Gill's Royal Appointment*

Gill's interests were in astrometry, precision measurement of positions in particular. Of course positional *changes* took him into the topic of stellar distances and stellar velocities and planetary and lunar positions into the realm of celestial mechanics. Nobody was better at astrometry than David Gill: rooting out every possible systematic error[3], using his considerable mechanical expertise to design minimum error systems and his watchmaking skills to build, modify or repair equipment. He was also a superb organiser, getting people to co-operate not by exercising formal authority but by rubbing off his own enthusiasm, leading by example and seeing problems from the point of view of others. I've chosen a few examples of his achievements.

*The size of the solar system*

There is one problem that David Gill pursued throughout his astronomical career. Maybe he was motived by George Biddell Airy repeating in the middle of the 19th century that it was the noblest problem in astronomy[4]. It was almost embarrassing to astronomers that they did not know precisely the size of the solar system. In mid-19th century, estimates using different techniques ranged by about 4%. The issue was not just one of knowledge of our immediate surroundings, though once the Earth-Sun distance is determined it also tells us through Kepler's 3rd law the distance to other planets, asteroids and comets and the size of the Sun and planets. Even the Admiralty were interested, for the Earth-Sun distance affects astronomical parameters then used to compile the ephemerides that underpin accurate marine navigation. Thirdly, the Earth-Sun distance is the basic 'metre stick' employed to measure distances to all other stars. Maybe Airy was right about its importance. Gill certainly thought so. Anyone wishing to measure the universe really needs to know the size of their reference length to a lot better accuracy than 4%.

Of course the Earth-Sun distance can't be measured directly as a distance, it must be deduced. The equivalent quantity is the equatorial parallax of the Sun – namely the angle subtended by the equatorial radius of the Earth at the centre of the Sun. We now know it is 8.794″ arc, not very big. Encke had deduced a value of 8.571″ arc earlier in the century from a detailed analysis of transit of Venus measurements and Le Verrier himself 8.95″ arc from an analysis of the perturbation of planetary orbits[5].

Gill's challenge was to start with the Paris metre, determine the best value for the equatorial radius of the Earth, measure the parallax to another solar system body and using Kepler's law deduce the parallax of the Sun, achieving a final answer accurate to a small fraction of a percent. It was a challenge that occupied Gill decades. He wasn't the only astronomer to spend years on this problem but his reward was that his answer became the accepted international value until at least the middle of the 20th century.

There are several methods of finding the solar parallax, the more indirect ones involve also knowing the speed of light, $c$, or both $c$ and the gravitational constant $G$. The solar parallax was why the Paris Observatory funded Foucault, Fizeau and Cornu to find ever more accurate values for $c$. Gill favoured a direct method proposed by Airy that had not been put to the test. He had first tried the method during an 1874 expedition to measure the transit of Venus. It was the method he would employ in the coming decades that came to be known as the *method of diurnal parallax*. Gill used it on Mars and a series of asteroids.

In brief the method involves using the rotation of the Earth to create the displacement of the two parallax observing stations, observations being made of the same astronomical body in the evening and pre-dawn by the same person using the same equipment. Using this technique, Gill finally defined the solar parallax as 8.80″ arc, accurate to almost 0.1%. In addition, his work determined the constant of aberration and the mass of the moon, both of his results also being taken in 1897 as the internationally accepted values.

*Heliometer*

Gill's weapon of choice in this and other precision measurements was the heliometer, a specialist telescope not seen these days. Gill became the master of the heliometer. He has described the instrument in the 11[th] edition of the Encyclopaedia Britannica[6]. Although the Admiralty's interest in stars was mainly confined to charting them, Gill's interests in astrometry extended much further. Gill's heliometer was smaller in size than the Königsberg instrument[7] Friedrich Bessel had used to first measure stellar parallax in the years prior to 1838. With it, however, Gill began a program at the Cape of Good Hope in 1881 to measure stellar parallaxes (i.e. star distances), assisted by the young visiting American astronomer William Elkin who was later to become Director of the Yale Observatory[8]. Before Elkin left, in 1882 Gill could say that they had measured more stellar distances than all the world's observatories put together[9]. Gill continued this program with a new 7 inch (180 mm) heliometer and it is interesting to read the words of that remarkable Cambridge astronomer Arthur Eddington who said in 1915 of Gill's work on stellar distances "*Whenever they have been put to the test, Gill's values have been confirmed. … when we compare the sizes of the instruments – the 40 inch telescope at Yerkes, or the 26 inch at Greenwich, with his 7 inch heliometer – we must marvel at the precision he could obtain.*"[10] We still marvel at Gill's precision. To give it some physical context, to measure the stellar parallax of even nearby stars to 10% requires a positional measuring accuracy of a few hundredths of an arc second. The diffraction disc for a perfect objective of 7 inch aperture is about 2.5 seconds of arc wide. Gill was determining accurate values to one percent of the width of the diffraction disk, and that includes the removal of systematic errors. All this was from an observatory that is almost at sea level, where the scintillation of stars is at its worst. He was truly operating right on the limits of the physically possible.

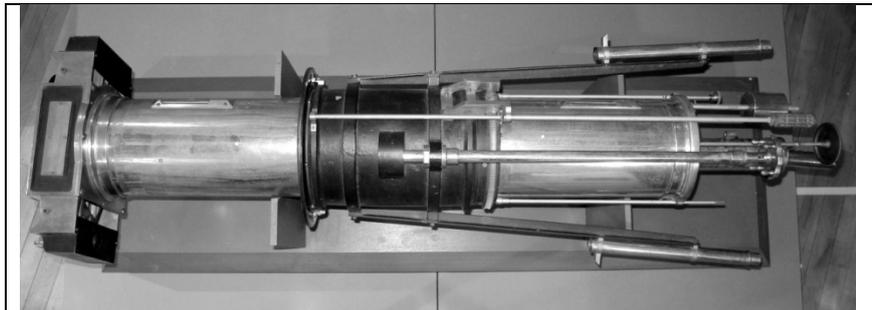

*Fig. 2 David Gill's 4 inch (102 mm) aperture heliometer on display in the Gill centenary exhibition at the Aberdeen City Maritime Museum.*

*Astrophotography and the CPD*

Gill pioneered the introduction of astrophotography into astrometry. His first venture into astrophotography was one that brought him to the notice of the serious astronomical community in Britain while he was still a watchmaker in Aberdeen. In 1869 he and a well-known local photographer took some rather good wet collodion plates of the Moon.

At the Cape Observatory in 1882, Gill and a local photographer used the new dry plate technique to photograph one of the 19[th] century's 'great comets'. Exposures of some half-hour or more showed detail in the comet tail that couldn't be seen clearly with the naked eye.

The photographs inspired Gill and other astronomers, not just with revelations about the comet but by the clarity of the background star fields. The time had come, Gill was sure, to use photography to extend the much used Bonn catalogue of mainly northern hemisphere stars right down to the southern celestial pole. The Bonn catalogue (the Bonner Durchmusterung – the Bonn 'selection', for it contained stars down to about 9[th] magnitude or a little fainter; positions given to the nearest 0.1 sec in right ascension and 0.1 arcmin in declination) had taken decades of eyepiece observing. Taking photographs was just the beginning but finding the accurate declination, right ascension and magnitude for every star to be catalogued, and it turned out there would be about 450,000, was a very big undertaking. The Dutch astronomer Jacobus Kapteyn came to the rescue, volunteering to build a machine to help scan the plates and make the catalogue. Gill set up a proper survey astrographic telescope at the Cape, with each plate covering about 5 degrees of field. That still implied well over 1000 good plates. It took 10 years to complete the first extensive photographic catalogue of stars but the Cape Photographic Durchmusterung became the southern hemisphere catalogue of choice and the photographic technique set the pattern for all future general catalogues..

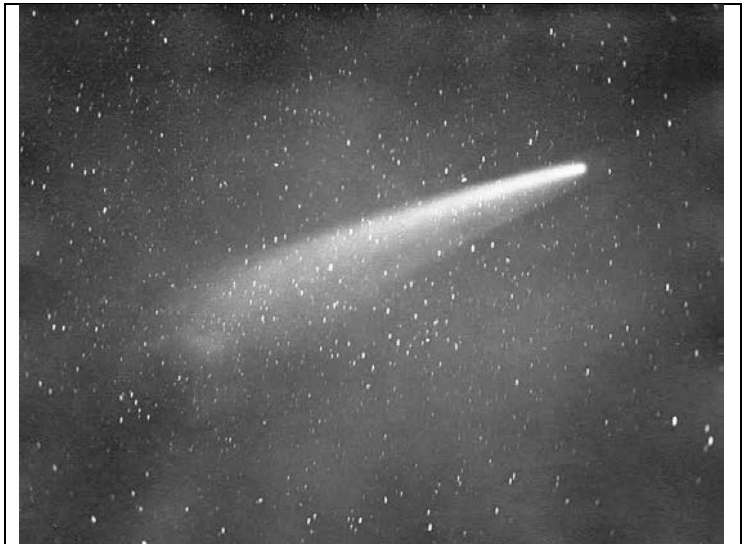

*Fig. 3 One of the photographs by David Gill of the Great Comet of 1882 that initiated the CPD catalogue and the Carte du Ciel project, courtesy Wikipedia.*

*The Carte du Ciel*

David Gill and Admiral Mouchez, The Director of the Paris Observatory who succeeded Le Verrier, were the founders of the *Carte du Ciel* project. The idea gelled in 1885. Both had experience in astrophotography. Photographic techniques were rapidly developing and both agreed that there was a "d*uty .. imposed on astronomers by the progress of celestial photography*", as Gill said[11], to use photography to comprehensively map the sky. Gill promoted the idea of taking the concept to the international community of astronomers via an international conference he proposed to be held in Paris.

The International Astrographic Congress of April 1887 was attended by 56 astronomers from 19 countries, with the net result that 18 observatories agreed to participate in creating a photographically based catalogue of the entire sky, pole to pole, containing stars to 11th magnitude and a map of stars to 14th magnitude[12]. Standard telescopes based on a design by the brothers Paul and Prosper Henry were to be used by all observatories[13]. Given the need to include each area of the sky on at least two plates to filter out emulsion blemishes and that shorter exposures are needed for the catalogue plates than for the map plates, over 40,000 photographs were needed, excluding failures due to drifting cloud, fogging by haze in the sky, emulsion malfunction, and so on. Gill himself called the project "the great monumental work of the century"[14] and anticipated it would take decades. It did.

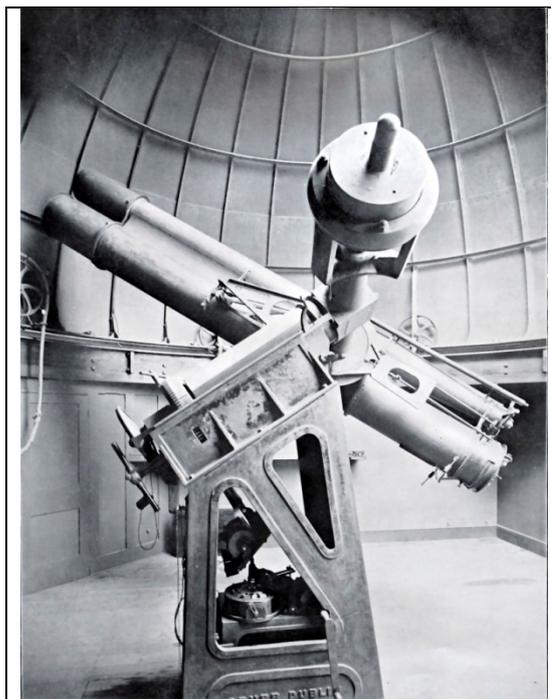

Fig. 4 The Cape Observatory's Carte du Ciel Astrographic telescope from David Gill "A History and Description of the Royal Observatory at the Cape of Good Hope" (London: HMSO, 1913).

The congress established the permanent International Astrographic Commission of which Mouchez was voted the executive President and Gill the Honorary President[15]. As Gill said, you don't need to bring astronomers from around the world together for over a week just to agree to photograph the sky but you do to make an all-sky catalogue with accurate stellar positions and magnitudes deduced for over 2 million stars; the map would contain some 40 million stars. Its purpose was not simply to record the geography of the sky but to provide a reference so that changes in time and changes in space, namely the variability and the motion of stars, could in future decades be found. The age-old ideal to record what was in the sky was part of the motivation but the motivation included the then modern notion of finding how star groups were moving.

Gill carried out the photography and plate measurement required for the Cape Observatory's assigned section of the sky, using a plate measurer of his own design made by Repsold of Hamburg. He provided general project technical guidance and regulations that had some intrinsic flexibility. Indeed through the rest of his career and into his retirement Gill continued to steer the project while four successive Directors of the Paris Observatory over the decades took the executive helm. The Permanent Commission spawned five sub-committees dealing with related astronomy. Gill provided some stability and continuity such a huge project required. In fact when he died in 1914 the project was not finished but Gill's vision of internationally co-operative astronomy and how to achieve it lived on, for the Astrographic Commission was joined by other bodies in 1919 to form the International Astronomical Union (IAU), the effective overseeing body for modern astronomy. The Commission's 4[th] President, Benjamin Baillaud, became the IAU's first President. It was, in short, Gill who planted one of the acorns from which the IAU grew.

Professor Kapteyn said in an encomium on Gill "*Outsiders who have seen him at work at these conferences may be under the impression that it was the geniality of his person, his infectious enthusiasm, and strong self-reliance that carried the day. But those who had followed matters closely would know how carefully he had studied every detail of the matter to be discussed, how long beforehand he had extensively corresponded with the most capable and interested persons, and how he brought many of them together a few days before the date of the congress, not only to arrange the program for the proceedings, but also to discuss informally all the main points. All through the congress, too, he would bring the ablest men together for these informal discussions.*"[16] This of course is how much modern science is done these days. Any astronomer who had not heard of David Gill before the Carte du Ciel project, would certainly have done so by the end of the century.

Gill himself said in 1907[17] "*A century hence this great work will have to be repeated, and then, if we of the present day have done our duty thoroughly, our successors will have the data for an infinitely more complete and thorough discussion of the motions of the sidereal system than any that can be attempted to-day.*" That is what is happening with the Hipparcos and Gaia galactic surveys, though of course astrometry came a long way from Gill's day even before this work. The Carte du Ciel project represented the limit of what astronomers could do in late 19[th] century and any science needs to develop on its limits. It seems unfair, as is sometimes done, to suggest that the Carte du Ciel project diverted astronomers from astrophysics since the physics we now know as necessary to understand the composition and evolution of stars, and the underpinning of cosmology, had not been established in Gill's lifetime.

*Some further achievements*

Gill's achievements in astrometry were impressive both to his contemporaries and in retrospect. He overhauled and extensively re-equipped the Cape Observatory, including installing a new, reversible, transit telescope designed with painstaking attention to fine detail and including many points of improvement over the Airy transit that the observatory had possessed. He produced a series of reference star catalogues derived from Cape Observatory transit measurements.

Gill is known also for his outstanding geodetic survey work. He established the best geodetic survey backbone in the southern hemisphere, stretching across South Africa and almost up to Lake Tanganyika. He also promoted the accurate survey of the 30° East longitude line as a means of obtaining a better figure for the shape of the Earth. The relevance of this to the determination of solar parallax concerned him all his life.

His extensive set of inner planetary observations gave a strong factual basis to solar system ephemerides. In the late 1880s and early 1890s Gill made more observations of Mercury than all other observatories put together and he sent these in correspondence to Simon Newcomb. Newcomb was refining the definitive model of solar system dynamics and Gill's observations helped Newcomb to arrive at the 43 seconds per century anomalous precession of Mercury's perihelion, a result that was crucial to Einstein.

*In retirement*

The Gill's retired to London a few months early due to overwork and some health problems. There he threw himself into the life of the scientific community. At various times he became President of the RAS, the BAAS, the IMarEst and the RDS (Research Defence Society). He was appointed British representative on the International Association of Geodesy and the International Committee for Weights and Measures. He was elected a council member of Royal Geographic Society and the first Honorary Member of the Astronomical and Astrophysical Society of America. One task he particularly enjoyed was that of delivering the Royal Institution Christmas lectures in 1907 on *Astronomy Old and New*. His home became the informal meeting point for visiting international astronomers.

It seemed that everyone wanted a slice of David Gill in retirement. To the end he was magnificent in his achievements and desirable in person. He died unexpectedly after a short illness in January 1914. According to his wishes, he was buried in the town where he had been born and spent most of the first 30 years of his life – Aberdeen, in NE Scotland. The centenary year of his death is an appropriate time to look back and assess David Gill's achievements. The next time you walk past the blue and white sign above a W H Smith bookshop, remember the man who gave Gill the opportunity to show what he could do for astronomy.

[1] James Lequeux *Le Verrier: Magnificent and Detestable Astronomer*, Springer, NY, 2013.

[2] Many personal details can be found in the biography written by his friend George Forbes *David Gill, Man and Astronomer*, John Murray, London, 1916. Gill was also a prolific correspondent and several thousand of his letters, or copies, survive in assorted archives. Individual references will not be quoted here.

[3] An example of the pains Gill took was quoted on his being presented with the Bruce Medal by the Astronomical Society of the Pacific where it was remarked that he had made 50,000 observations to determine the division errors of his heliometer. George C. Pardee *Address of the retiring President of the Society, in Awarding the Bruce Medal to H. M. Astronomer, Dr David Gill*, Publications of the Astronomical Society of the Pacific, vol. XII, no. 70, pp 49 – 55, 1st April 1900.

[4] This remark has been widely attributed to Airy, though as reported in Monthly Notices of the Royal Astronomical Society 17 (7): 208, May 8, 1857 he said "The measure of the Sun's distance has always been considered the noblest problem in astronomy".

[5] A useful historical summary is given by Agnes M. Clerk *A popular history of astronomy during the nineteenth century*, Chapter VI, Adam & Charles Black, London, 1st edition 1885, 4th edition 1902.

[6] [David Gill] article *Heliometer* in 'The Encyclopaedia Britannica', vol. XIII, pp 224 – 230, The Encyclopaedia Britannica Co., NY, 1910.

[7] Fraunhofer's Königsberg heliometer had an object of diameter 157 mm, Gill's instrument by Repsold although of superior design had an objective of 102 mm diameter.

[8] Elkin was almost certainly attracted to Yale and vice-versa because they had acquired a Repsold heliometer in 1882. He embarked on an extensive series of parallax measurements at Yale that was the most comprehensive of the late 19th century.

[9] Letter to E. B. Knobel, 17th April 1882 "The whole of Elkin 's work combined with mine is greater in extent than all the existing parallax determinations put together." Quoted in op. cit. 2, p 147.

[10] H. E. Krumm *Gill's work on stellar parallax with the heliometer* Astronomical Society of South Africa Monthly Notes, vol. II, no. 10, 1943, quoted on pp 88 – 89.

[11] Ileana Chinnici *La Carte du Ciel: Correspondence inédite conservée dans les archives de* l'Observatoire *de Paris, p86* (Observatoire de Paris, Paris & Osservatorio Astronomico di Palermo, Palermo, 1999).

[12] 14th magnitude is about the faintest star that can just be seen in the eyepiece of a telescope of 300 mm diameter objective, the largest in reasonably common use at the time.

[13] The Henry's prototype telescope copied for the project had a 350 mm objective and focal length of 3.44 m, making 1 minute of arc extend 1 mm on the photographic plate. Measuring coordinates to better than an arc second was possible. An area of sky 2 degrees square could be covered on one photographic plate.

[14] Ileana Chinnici, op. cit. 11, p 112.

[15] Forbes, op. cit. 2.

[16] J. C Kapteyn "Sir David Gill" The Astrophysical Journal, vol. XL, no. 2, Sept. 1914, pp 161 – 172.

[17] David Gill, Presidential address in "Report of the seventy-seventh meeting of the British Association for the Advancement of Science Leicester 31 July – 7 August 1907", John Murray, London 1908.